\documentclass[conference]{IEEEtran}
\IEEEoverridecommandlockouts
\usepackage{cite}
\usepackage{amsmath,amssymb,amsfonts}
\usepackage{algorithmic}
\usepackage{algorithm}
\usepackage{graphicx}
\usepackage{textcomp}
\usepackage{xcolor}
\usepackage{url}
\usepackage{booktabs}
\usepackage{bm}
\usepackage{siunitx}
\def\BibTeX{{\rm B\kern-.05em{\sc i\kern-.025em b}\kern-.08em
    T\kern-.1667em\lower.7ex\hbox{E}\kern-.125emX}}

\newtheorem{lemma}{Lemma}

\begin{document}

\title{CRLB and Parameter Estimation for OFDM-ISAC with Non-Uniform Sparse Resource Allocation}

\author{
    \IEEEauthorblockN{{Wenjie~Zhang\IEEEauthorrefmark{1}, Qianglong~Dai\IEEEauthorrefmark{1}, Xiaoli~Xu\IEEEauthorrefmark{1}, Ruoguang~Li\IEEEauthorrefmark{2}, and Yong~Zeng\IEEEauthorrefmark{1}\IEEEauthorrefmark{3}}
    \IEEEauthorblockA{\IEEEauthorrefmark{1} National Mobile Communications Research Laboratory, Southeast University, Nanjing 211100, China}
    \IEEEauthorblockA{\IEEEauthorrefmark{2} School of Information Science and Engineering, Hohai University, Changzhou 213200, China}
       \IEEEauthorblockA{\IEEEauthorrefmark{3} Purple Mountain Laboratories, Nanjing 211100, China}
    \IEEEauthorblockA{ \{wenjiezhang, qldai, xiaolixu,\}@seu.edu.cn, \{ruoguangli\}@hhu.edu.cn,\{ze0003ng\}@e.ntu.edu.sg}
    \thanks{This work was supported in part by the National Natural Science Foundation of China under Grants 62301157 and 62571116, in part by the Natural Science Foundation of Jiangsu Province under Grant BK20230823, and in part by the Fundamental Research Funds for the Central Universities under Grants 2242022k60004 and 3204002004A2.}
}
}

\maketitle

\begin{abstract}
Integrated sensing and communication (ISAC) holds great promise in expanding the applications of wireless communication networks.
However, in current communication-centric systems, the time-frequency resources available for sensing may be limited, and also usually non-uniformly and sparsely distributed across the time-frequency domain.
Such a non-uniformity destroys the ``thumbtack-shaped" ambiguity function of the orthogonal frequency division multiplexing (OFDM) waveform, leading to degraded sensing performance.
To this end, this paper explores the parameter estimation algorithm for OFDM-ISAC systems with non-uniform sparse resource allocation.
Specifically, for the single target case, we derive the closed-form Cram{\'{e}}r-Rao lower bound (CRLB) for parameter estimation as a function of resource indices. 
Furthermore, we show that simply filling unused resource locations with zeros and applying the classic periodogram estimation is equivalent to maximum likelihood (ML) estimation, which is asymptotically optimal.
For the multi-target case, we generate a virtual resource using the autocorrelation function of the original signal, which exhibits a significantly larger virtual bandwidth compared to the original signal, at the cost of higher peak-to-sidelobe ratio (PSLR).
Simulation results demonstrate that the proposed approach outperforms the conventional periodogram method for non-uniform sparse resource allocation.
\end{abstract}

\section{Introduction}
Integrated sensing and communication (ISAC) has been identified as one of the key usage scenarios for sixth-generation (6G) mobile networks \cite{b41}. 
It enables expansion of current communication-centric applications without additional cost of deploying standalone sensing platforms. 
The enhancement in sensing capabilities within communication systems can be attributed to the increase in operating frequency and the broadening of the bandwidth, which provide sparse propagation paths and sensing information about the environment \cite{3GPP_ISAC}. 
Moreover, orthogonal frequency division multiplexing (OFDM) has been the dominant waveform in fifth-generation (5G) mobile communication and still envisioned as a candidate waveform in 6G. 
This waveform not only excels in data transmission but also possesses several properties that make it suitable for sensing services, such as flexible time-frequency resource allocation, the array-manifold-like structure in subcarrier and symbol domain, and the ``thumbtack-shaped" ambiguity function.

Most existing works on ISAC have assumed continuous time-frequency resources for sensing, allowing conventional signal processing techniques for OFDM radar \cite{Braun2014} to be directly applied \cite{cui2021integrating,dai2025tutorial}. 
However, when sensing capabilities are integrated into an existing communication system, their impact on communication services should be minimized, which in turn imposes stringent constraints on sensing resource allocation. 
On the other hand, to ensure accurate parameter estimation and high resolution performance, a sufficient number of OFDM symbols and successive subcarriers within the coherent processing interval (CPI) are necessary. 
To achieve this, multiple orthogonal resource elements (REs) may be aggregated for sensing. However, in the practical mobile communication system, the resources available for sensing are limited, since a considerable proportion of them have been occupied by the dedicated signal such as physical downlink shared channel (PDSCH), physical downlink control signal (PDCCH), etc. 
Besides, if sensing services are provided for multiple users, the total time-frequency resources need to be split among them.
Non-uniform resource allocation serves as an effective method to addressing the aforementioned issues, which can enhance the flexibility, while the cost is the elevated sidelobes in the delay-doppler periodogram \cite{10636690,liu2022integrated}. 
In fact, non-uniform sparse resource allocation has attracted considerable research interest in sensing waveform design and array design. 
Non-uniform and sparse resource allocations with suppressed sidelobes can be constructed using strategies such as random search, co-prime\cite{10570265}, nested \cite{5456168}, or minimum redundancy \cite{1139138}.

Instead of focusing on sparse resource allocation strategies, this paper investigates the parameter estimation at the receiver side when the allocated resources of sensing signal are sparse and random.
The motivation for this paper is twofold. 
First, when available sensing resources are limited, all resources must be utilized to ensure acceptable sensing performance, leaving no room for further selection. 
Second, the analytical performance expressions of the optimal sensing algorithms can provide valuable guidance for resource allocation at the transmitter.

In this paper, we first present the likelihood function of the target parameter given the received information on the selected non-uniform sparse resource elements and formulate the parameter estimation problem to maximize this likelihood function. 
For single target sensing, the maximum likelihood (ML) estimation is derived by taking into account the random phase shift. 
It is shown that in this case, the ML estimation can be effectively computed by filling the unused resource allocations with zeros and then applying Discrete Fourier Transform (DFT) to obtain the parameter estimation.
Additionally, we derive  the Cram{\'{e}}r-Rao lower bound (CRLB) as a function of resource indices, which indicates that improved estimation performance can be achieved if the random and sparse resource indices cover a large virtual bandwidth.
This finding aligns with expectations and supports existing design policies such as those used in co-prime and nested methods.
Next, for general scenarios with multiple targets, we point out that the optimal estimation algorithm resembles the classic spectrum estimation problem \cite{Nohrden:SpectralEstimation:1995}, which has a very complex CRLB expression \cite{655430}.
We recommend using the autocorrelation method to obtain the sensing signal on the virtual resource, which exhibits a significantly larger virtual bandwidth compared to the original signal. 
The remaining holes in the virtual resource elements are simply filled with zeros, and then DFT is adopted to estimate the target parameters from the sensing signal on the virtual resource.
Simulation results verify the insights obtained from the analytical expressions and demonstrate the performance of the autocorrelation method in target parameter estimation with random non-uniform sparse resource allocations.

\section{System model}\label{sec:sysmodel}
\begin{figure}[h]
    \centering
    \includegraphics[scale=0.23]{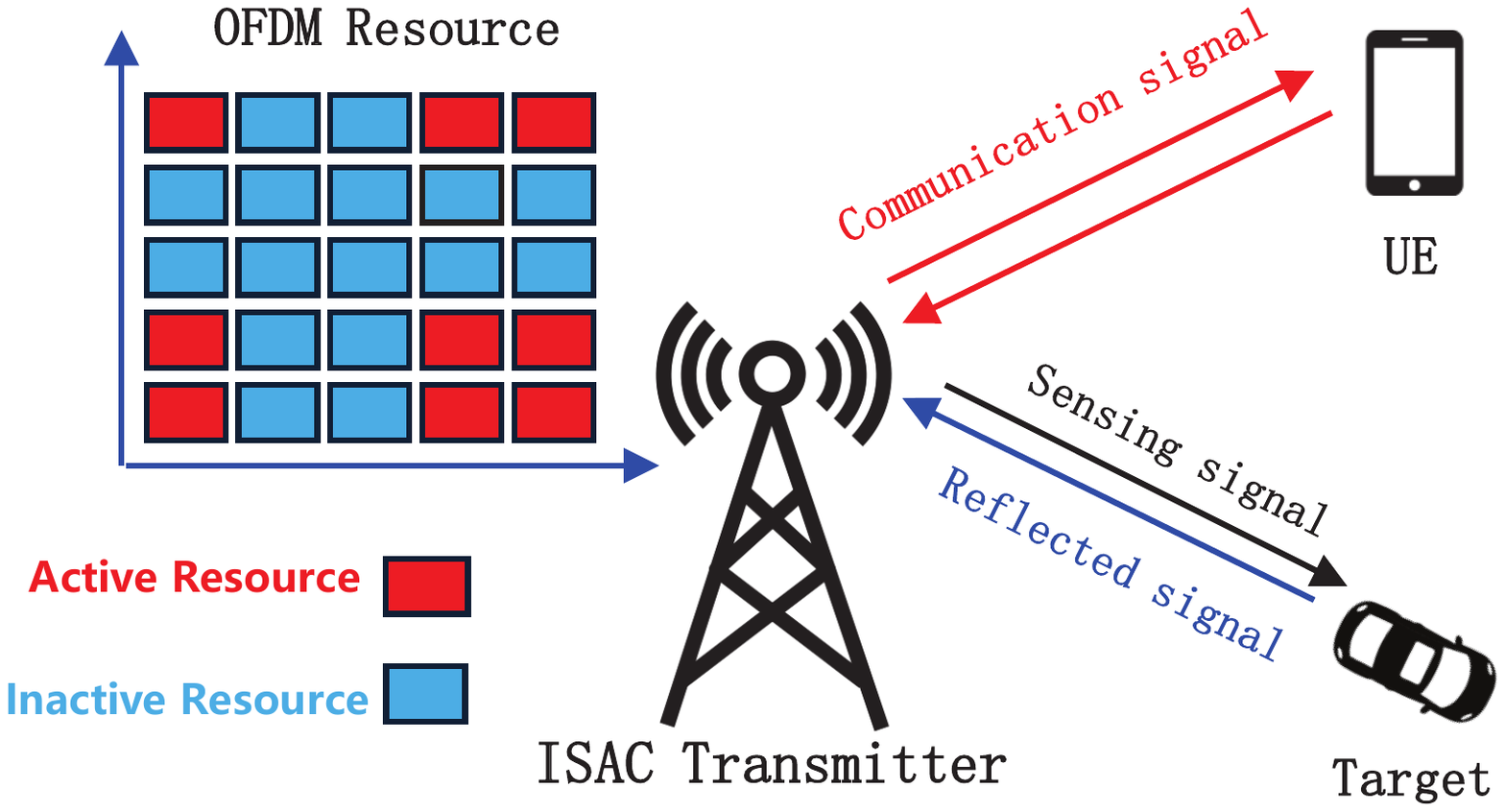}
    \caption{OFDM-ISAC system with non-uniform sparse resource allocation. }
    \label{F:model}
\end{figure}

As shown in Fig.~\ref{F:model}, we consider a mono-static OFDM-ISAC system with non-uniform sparse resource allocation. 
Specifically, denote the system bandwidth by $B$, which is divided into subcarriers with inter-carrier spacing $\Delta f$. 
The number of subcarriers is $N=B/\Delta f$.
And the OFDM symbol duration is $T=1/\Delta f$.
Furthermore, to tackle the inter-symbol-interference (ISI) caused by the multipath channel,a cyclic prefix (CP) of length $T_{cp}$ is inserted, rendering the effective OFDM symbol duration $T_s=T+T_{cp}$. 
Denote the coherent processing interval (CPI) by $T_{\mathrm{total}}$, and the number of OFDM symbols for sensing is $M=T_{\mathrm{total}}/T_s$.
A resource element (RE)  is defined as one subcarrier in frequency domain and one OFDM symbol in time domain, indicated by the block in Fig.~\ref{F:model}.
Due to the coexistence of communication and sensing, we assume that only a subset of RE could be used for sensing.
For each symbol, the $N$ subcarriers are denoted as a set $\mathcal{N}=\{0,1,2,...,N-1\}$, and the set of subcarriers used for sensing in the $m$th OFDM symbol is denoted by $\mathcal{N}_m$, where $\mathcal{N}_m\subseteq \mathcal{N}$ and $m=0,1,...,M-1$.
If $\mathcal{N}_m=\emptyset$, the $m$th symbol is not used for sensing at all.

\subsection{Signal Model}
With non-uniform transmission, the OFDM signal sent by the ISAC transmitter for sensing is given by
\begin{align}
x(t)= \sum_{m=0}^{M-1} \sum_{k\in\mathcal{N}_m} d_{k,m} e^{j 2 \pi k \Delta f (t - m T_s)}
\mathrm{rect}\left( \frac{t - m T_s}{T_s} \right)\label{xt}.
\end{align}
where $d_{k,m}$ is the data carried by the $k$th subcarrier of the $m$th OFDM symbol, and $\mathrm{rect}(\cdot)$ is the rectangular function. 
If the selected resource is used purely for sensing, we can set $d_{km}=1$ to maximize the sensing performance. 
If a random data symbol is carried, $d_{km}$ is usually removed by dividing it from the received signal, since it is totally known at the ISAC transmitter \cite{Braun2014}.

Assume there are $L$ targets and the multipath time-variant channel for sensing can be modeled as
\begin{align}
h(t,\tau)=\sum_{l=1}^{L}\alpha_l\delta(\tau-\tau_l)e^{j2\pi f_{D,l}t}.
\end{align}
where $\delta[\cdot]$ is the unit impulse function, $\alpha_l$ is the channel gain, $\tau_l$ is the delay and $f_{D,l}$ is the Doppler shift of the $l$th path. The delay and  Doppler shift are related with the target distance $d_l$ and radial velocity $v_l$ as
\begin{align}
\tau_l=\frac{2d_l}{c},\quad
f_{D,l}=\frac{2v_lf_c}{c}.\label{eq:tau_fD}
\end{align}
where $c$ is the light speed and $f_c$ is the carrier frequency.

Thus, the reflected signal received by the ISAC BS can be expressed as
 \begin{align}
 y(t)&=x(t)*h(t,\tau)+w(t)\nonumber\\
 &=\sum_{l=1}^{L}\alpha_l x(t-\tau_l) e^{j2\pi f_{D,l}t}+w(t). \label{eq:ytch}
 \end{align}
where $w(t)$ is the additive white Gaussian noise (AWGN) with power spectrum $\mathcal{W}(f)=N_0/2$.

\subsection{Target Estimation}
For target sensing, the ISAC BS wants to estimate the target parameters $\theta=\{\tau_l,f_{D,l},l=1,...,L\}$ from the received signal $y(t)$. 
After self-interference cancel and CP removal \cite{9724187}, sample the received signal with frequency $B=N/T$ and the discrete samples for the $m$th OFDM symbol can be written as
\begin{align}
y_m[n]=\sum_{l=1}^{L}&\alpha_le^{j2\pi  f_{D,l}mT_s}\sum_{k\in\mathcal{N}_m}e^{j\frac{2\pi}{N}k(n-\tau_lB)}+w_m[n],\nonumber\\
& n=0,1,...,N-1. \label{eq:ymn}
\end{align}
Next, taking $N$-point FFT on $y_m[n]$ renders
\begin{align}
&Y_m[n]=\sum_{l=1}^{L}\alpha_le^{j2\pi  f_{D,l}mT_s}e^{-j2\pi n\Delta f\tau_l}+W_m[n]. \label{eq:Y_m}
\end{align}
where $W_m[n]$ is the noise sample with variance $N_0/2$.

With decoupling and symmetry of delay and Doppler estimation, this paper takes delay estimation as an example for analysis. The method and results can be readily extended to the Doppler estimation. 
Furthermore, the power of channel gain $\alpha_l$ is related to the radar cross section (RCS) of the target and given by
\begin{align}
|\alpha_l|^2=\frac{\kappa_l\lambda^2}{(4\pi)^3d_l^4}G_TG_RP_T.
\end{align}
 where $\kappa_l$ is the RCS of the $l$th target, $d_l$ is the distance of the $l$th target, $\lambda$ is the wavelength, $G_T$ and $G_R$ are transmit and receive gain.
 For notational convenience, we denote $\alpha_l=A_le^{j\phi_l}$, where $A_l=|\alpha_l|$ is the amplitude, and $\phi_l$ is the random phase. 
 Then, the ML estimation is obtained by
\begin{align}
\textbf{P1:  } \underset{\{\tau_l,l=1,...,L\}}{\max}\Pr(Y_m[i]|\{A_l,\phi_l,\tau_l,\forall l\}). \label{eq:P1}
\end{align}
where
\begin{small}
\begin{align}
&\Pr(Y_m[i]|\{A_l,\phi_l,\tau_l,\forall l\})=\left(\frac{1}{\sqrt{\pi N_0}}\right)^{\sum_{m=0}^{M-1}|\mathcal{N}_m|} \nonumber\\
&\exp\left(\sum_{m=0}^{M-1}\frac{\sum_{i\in\mathcal{N}_m}\|Y_m[i]-\sum_{l=1}^{L}A_l e^{-j(2\pi i\Delta f\tau_l-\phi_l)}\|^2}{N_0}\right).
\end{align}
\end{small}

Drop the irrelevant constant and consider the monotonically increasing property of exponential function, \textbf{P1} can be reduced to \textbf{P2}, given by
\begin{small}
\begin{align}
\textbf{P2:  } \underset{\{\tau_l,l=1,...,L\}}{\min}\sum_{m=1}^{M}\sum_{i\in\mathcal{N}_m}\|Y_m[i]-\sum_{l=1}^{L}A_le^{-j(2\pi i\Delta f\tau_l-\phi_l)}\|^2. \label{eq:P2}
\end{align}
\end{small}

If the sensing resource is continuous, i.e., $\mathcal{N}_m=\mathcal{N},\forall m$, the unknown parameter $\{\tau_l,l=1,...,L\}$ can be estimated using periodogram, by taking FFT on the subcarrier indices. However, when the OFDM resource is non-uniform, the optimal sensing algorithm is unclear. This paper derives the CRLB of the parameter estimation shown in \textbf{P2} when there is a single target, and we propose an efficient estimation algorithm for the general case.

\section{CRLB and Parameter Estimation For a Single Target}
When there is only one target, \textbf{P2} reduces to
\begin{align}
\textbf{P2(a):  } \underset{{\tau}}{\min}\sum_{m=1}^{M}\sum_{i\in\mathcal{N}}\|Y_m[i]-Ae^{-j(2\pi i\Delta f\tau-\phi)}\|^2. \label{eq:P}
\end{align}

The conditional probability density function (PDF) of the received samples $\{Y_m[i],i\in \mathcal{N}\}$ given $\tau$ and $\phi$ can be further simplified to
\begin{align}
&\Pr(\{Y_m[i]\}|\tau,\phi)\nonumber\\
&=C\exp\left(\frac{2A}{N_0}\sum_{m=1}^{M-1}\sum_{i\in\mathcal{N}}\Re\left\{Y_m[i]e^{j(2\pi i\Delta f\tau-\phi)}\right\}\right).  \label{eq:likelihood}
\end{align}
where $C$ is a constant independent of the parameter estimation and $\Re\{x\}$ denote the real part of $x$. Take the log of \eqref{eq:likelihood} and drop the constant, we obtain the log-likelihood function as
\begin{align}
\Lambda_L(\tau,\phi)=\frac{2A}{N_0}\sum_{m=1}^{M-1}\sum_{i\in\mathcal{N}}\Re\left\{ Y_m[i]e^{j(2\pi i\Delta f\tau-\phi)}\right\}. \label{eq:Loglikelihood}
\end{align}

Take the expectation of second order derivatives of $\Lambda_L(\tau,\phi)$, we can obtain the Fisher's information matrix (FIM) as
\begin{align}
&\mathbf{I}
=\left[\begin{array}{cc}
-\mathbb{E}\left[\frac{\partial^2 \Lambda_L(\tau,\phi)}{\partial \tau^2}\right] & -\mathbb{E}\left[\frac{\partial^2 \Lambda_L(\tau,\phi)}{\partial \tau\partial \phi}\right]\\
-\mathbb{E}\left[\frac{\partial^2 \Lambda_L(\tau,\phi)}{\partial \phi \partial \tau} \right]&
-\mathbb{E}\left[\frac{\partial^2 \Lambda_L(\tau,\phi)}{\partial \phi^2}\right]
\end{array}\right]\nonumber\\
&=\frac{2A^2}{N_0}\sum_{m=0}^{M-1}\left[
   \begin{array}{cc}
     \sum_{i\in\mathcal{N}}(2\pi\Delta fi)^2 & \sum_{i\in\mathcal{N}}2\pi\Delta fi \\
     \sum_{i\in\mathcal{N}}2\pi\Delta fi & |\mathcal{N}| \\
   \end{array}
 \right].
 \label{eq:FIM}
\end{align}

Based on \eqref{eq:FIM}, we can obtain the CRLB for the unbiased estimation variance of $\tau$ as
\begin{align}
\sigma_{\tau}^2\geq \frac{N_0}{2A^2}\frac{\sum_{m=0}^{M-1}|\mathcal{N}|}{(\sum_{m=0}^{M-1}|\mathcal{N}|)\cdot g_1(\{\mathcal{N}\})-g_2(\{\mathcal{N}\})},
\end{align}
where $g_1(\{\mathcal{N}\})=\sum_{m=0}^{M-1}\sum_{i\in\mathcal{N}}(2\pi\Delta fi)^2$ and $g_2(\{\mathcal{N}\})=(\sum_{m=0}^{M-1}\sum_{i\in\mathcal{N}}2\pi\Delta fi)^2$.

Under the special case when the same set of subcarrier is used for all $M$ OFDM symbols, i.e., $\mathcal{N}_{m}=\mathcal{N}_0,\forall m$, we have
\begin{align}
\sigma_{\tau}^2\geq \frac{N_0}{2MA^2}\frac{|\mathcal{N}_0|}{|\mathcal{N}_0| \sum_{i\in\mathcal{N}_0}(2\pi\Delta fi)^2-(\sum_{i\in\mathcal{N}_0}2\pi\Delta fi)^2}.
\label{eq:CRLB}
\end{align}

Next, we consider the optimal estimation algorithm for $\tau$ in \textbf{P2(a)}. For non-coherent estimation, we assume that $\phi$ is uniformly distributed in $[0,2\pi)$ and follow the analysis in \cite{600934}, the ML estimate is given by
\begin{align}
\hat{\tau}=\underset{\tau}{\arg\max}\left|\sum_{m=0}^{M-1}\sum_{i\in\mathcal{N}}Y_m[i]e^{j2\pi i\Delta f\tau}\right|.\label{eq:ML}
\end{align}

On the other hand, treating the phase as the fading coefficient of the observations, the ML estimation was shown to be equivalent to calculating the discrete-time Fourier transform (DTFT) of the correlation function of the received signal \cite{600934}. 

For multiple targets scenarios, the CRLB of the targets' delays estimation is extremely complex and the optimal ML estimation is challenging to achieve due to the strong coupling between multiple unknown parameters ${A_l, \phi_l, l=1,\cdots,L}$ and inter-target interference \cite{10994424}. 


\section{Parameter Estimation for Multiple Targets}
For continuous resource allocation,  the DFT algorithm can be directly used to obtain the delay estimation priodogram with low side-lobes, while the achievable resolution is low. 
However, for sparse resource allocation schemes, resolution approaching the full system bandwidth can be achieved using partial bandwidth, but the delay estimation periodogram may exhibit undesirable high sidelobes, where weak targets can be easily masked by strong ones. In the literature, the commonly adopted estimation algorithms for non-uniform sampling \cite{Nohrden:SpectralEstimation:1995} include
\begin{itemize}
\item{{\it Throw-away method}: This method discards the information received in some resource, so that the remaining one forms a good pattern, e.g., uniform sparse. }
\item{{\it Autocorrelation method:} This method takes the expected value of correlation between two selected resources with the same frequency gap, and hence obtains sensing signal on the virtual resource with higher resolution.   }
\item{{\it Interpolation method:} This method uses some prior information of the target delay to obtain a better interpolation for the information at an unused resource location. }
\item{{\it Filter bank method}: This method keeps the original received information on the selected resource, and generates the interpolated information on the unused resource using the filter bank.}
\end{itemize}

Among the above methods, the throw-away method has good performance only when the selected resource occupies a large portion of the total resource, which is not the case considered in this paper. On the other hand, the interpolation method has requirement on prior information, which is not readily obtainable, and the filter bank method has high implementation complexity. To this end, we adopt the autocorrelation method, aiming to create larger virtual resources based on the limited selected resources available for sensing in ISAC, which is also computational-efficient.

Taking frequency resource sparse allocation as an example, denote $\tilde{N}$ as the allocated sparse subcarrier subset of the continuous subcarrier resource $\mathcal{N}$, \eqref{eq:Y_m}  can be reduced to
\begin{small}
    \begin{align}
    Y_m[n]=\begin{cases}
        \sum_{l=1}^{L}\alpha_le^{j2\pi  f_{D,l}mT_s}e^{-j2\pi n\Delta f\tau_l}+W_m[n],& n\in\tilde{N}\\
        0,&else
    \end{cases}
\end{align}
\end{small}

For simplicity, the following analysis considers consecutive OFDM symbols with the same $\tilde{N}$, and the signal over all OFDM symbols at the same subcarrier position can be expressed as
\begin{align}
\label{matrix Y}
\bm{Y}_m=[Y_m[0],\,\ldots \,,Y_m[N-1]]^T \in \mathbb{C}^{N}.
\end{align}

For the classical DFT-based method, the delay parameters  $\{\tau_l,\forall l\}$ are estimated from the delay periodogram obtained by applying an inverse discrete Fourier transform (IFFT) to the sensing signal. When multiple OFDM symbols are superposed in the time domain,  the sum of contributions from all  OFDM symbols  can be used for joint delay estimation, given by
\begin{small}
\begin{align}
    \frac{1}{M} \sum_{m}^{M-1}&\mathcal{F}^{-1} \{\bm{Y}_m\}_q=\frac{1}{MN} \sum_{m=0}^{M-1}\sum_{n\in \tilde{N}}(Y_m[n])e^{j 2 \pi \frac{qn}{N}}
     \label{eq:simply IDFT}.
\end{align} 
\end{small}
where $\mathcal{F}^{-1}\{\cdot\}$ denotes the IFFT operator. From \eqref{eq:simply IDFT}, it is observed that the peaks associated with the target delay parameters are at the indices of $q_l=\frac{2N\Delta fd_l}{c}$. Sidelobes exits at random locations when $\tilde{N}$ is only a sparse subset of the continuous subcarriers. 

The computation of autocorrelation may generate a virtual signal with significantly higher density, thereby reducing the sidelobe levels. Let $S$ denote the set of the virtual resource. The sensing signal on the virtual resource of position $s$ can be approximated as
\begin{align}
\tilde{R}_m[s]
&= \frac{1}{c[s]}\sum_{i=0}^{N-1} Y_m[i]\,Y_m^*[i+s] \nonumber\\
&= \tilde{R}_{m,\tau,s}[l;p]+\tilde{R}_{m,c,s}[l;p]+\tilde{R}_{m,w,s}[l;p]. 
\end{align}
where the entry $\tilde{R}_{m,t,s}[l;p]$ is
\begin{small}
\begin{align}
\label{Auto}
\tilde{R}_{m,\tau,s}[l;p]=
&\begin{cases}
    \sum\limits_{l}^{L}|a_l|^2e^{-j2\pi\Delta f s\tau_l},l=p.\\
    \frac{1}{c[s]}\sum\limits_{i\in \tilde{N}}\sum\limits_{l,p}\sum\limits_{l\neq p}a_la_p^*e^{-j2\pi \Delta fs\tau_p}  \delta_{l,p}^{D,m}\delta_{,l,p}^{\tau,i},l\neq p.
\end{cases}
\end{align}
\end{small}
and $ \delta_{l,p}^{D,m}=e^{j2\pi mT_s(f_{D,l}-f_{D,p})}$, $\delta_{l,p}^{\tau,i}=e^{-j2\pi \Delta f i(\tau_l-\tau_p)}$. $\tilde{R}_{m,c,s}[l;p]$ is the signal–noise cross term. $\tilde{R}_{m,w,s}[l;p]$ is the noise cross term. 

Without loss of generality, we assume that the allocated resource includes the first and last resource. For a given subcarrier index set $\tilde{\mathcal{N}}$, following the implementation of \eqref{eq:virtual signal indices}, the virtual resource indices are defined as
\begin{align}
    \mathcal{P}&=\{  n_i - n_j  | \ \forall n_i, n_j \in \tilde{\mathcal{N}} \}.
    \label{eq:virtual signal indices}
\end{align}

The sensing signal on the virtual resource of the $mth$ OFDM symbol can be expressed as $\bm{\tilde{R}}_m=[R_m[-N+1],\cdots,R_m[0],\cdots ,R_m[N-1]]^T$, $|\bm{\tilde{R}}|=|\mathcal{P}|$. 

\begin{lemma}
If we randomly select Na subcarriers from N continuous subcarriers, the probability is
\begin{align}
\label{hole pr}
Pr(d, N_a, N) = \frac{ \sum_{m}\prod_{r=0}^{s-1}N_r(m_r)}{\binom{N-2}{N_a-2}}.
\end{align}
where $\boldsymbol{m} = \{m_0, m_1, \dots, m_{d-1}\}$ is a set of non-negative integers satisfying $\sum_{r=0}^{d-1} m_r = N_a$.The proof is omitted here for brevity.
\end{lemma}

 \begin{figure}[htbp]
    \centering
    \includegraphics[scale=0.23]{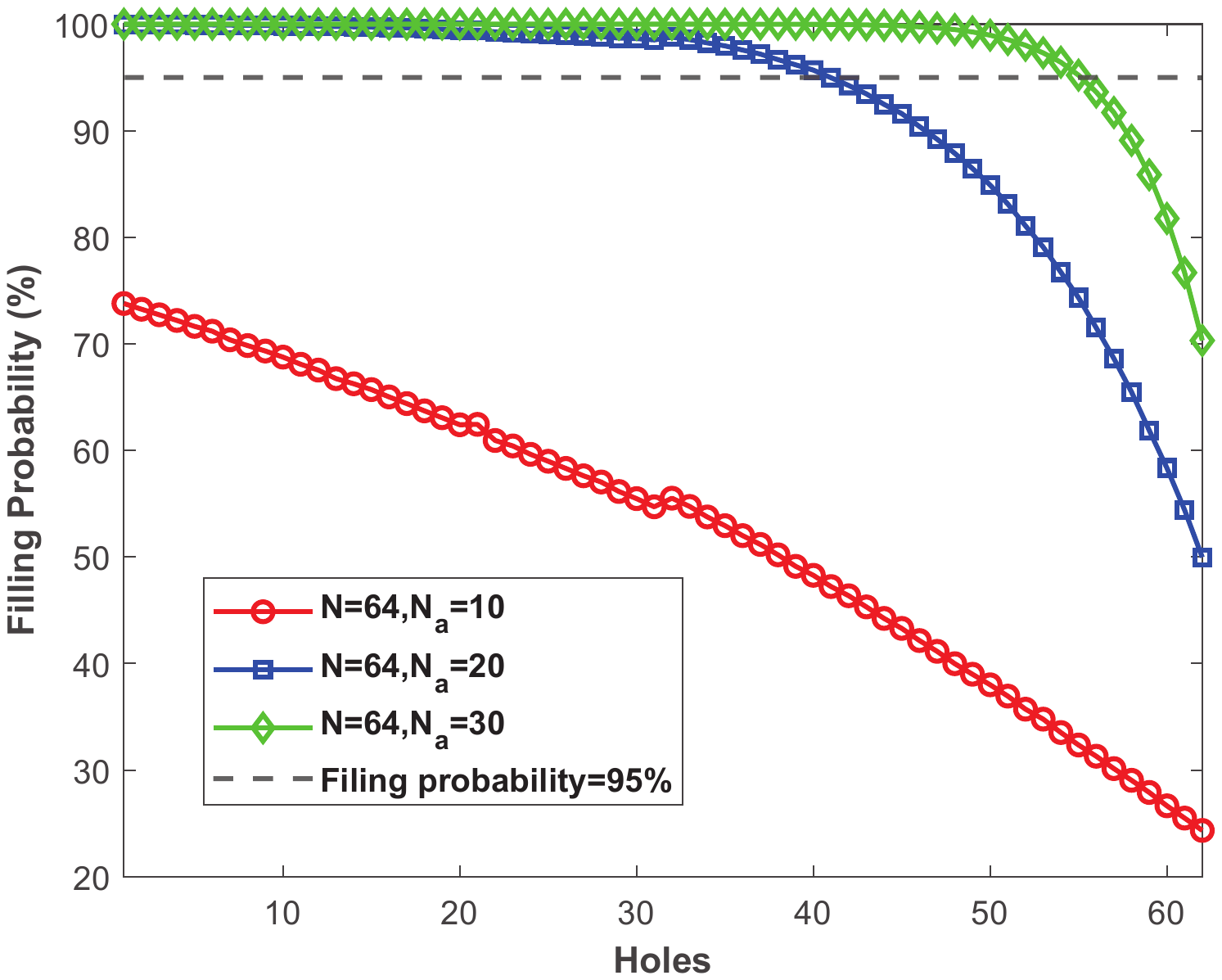}
    \caption{Probability for filling a hole in virtual signal based on randomly generated sparse resource.}
    \label{hole}
\end{figure} 

If a particular index is not filled in the virtual signal, it is deemed as a ``hole" of the virtual signal. Fig. \ref{hole} illustrates the probability that a particular hole is filled in the virtual signal, based on \eqref{hole pr}. As the allocated resources increase, the probability of obtaining virtual resource positions through differencing increases. With 95\% as the threshold, the total number of virtual resource positions that can be obtained is generally greater than the number of original resources $|\mathcal{P}|>|\tilde{\mathcal{N}}|$. 
\addtolength{\topmargin}{0.07in}
Then the sensing signal on the virtual resource over $M$ OFDM symbols is $\bm{\tilde{R}}=[\bm{\tilde{R}}_0,\bm{\tilde{R}}_1,\cdots ,\bm{\tilde{R}}_{M-1}]$.
The cross term $\bm{\tilde{R}}_{c,s}$ and $\bm{\tilde{R}}_{w,s}$ can be suppressed through coherent accumulation across OFDM symbols in the same CPI due to the characteristics of complex Gaussian white noise, given by
\begin{align}
\label{eq:26}
    E\,[ \bm{\tilde{R}}]&=E\,[\bm{\tilde{R}}_{\tau,s}]+E\,[\bm{\tilde{R}}_{c,s}]+E\,[\bm{\tilde{R}}_{w,s}]\\ \nonumber
    &= \sum\limits_{l}^{L}|a_l|^2e^{-j2\pi\Delta f s\tau_l}
\end{align}

\begin{figure}[htbp]
    \centering
    \includegraphics[scale=0.23]{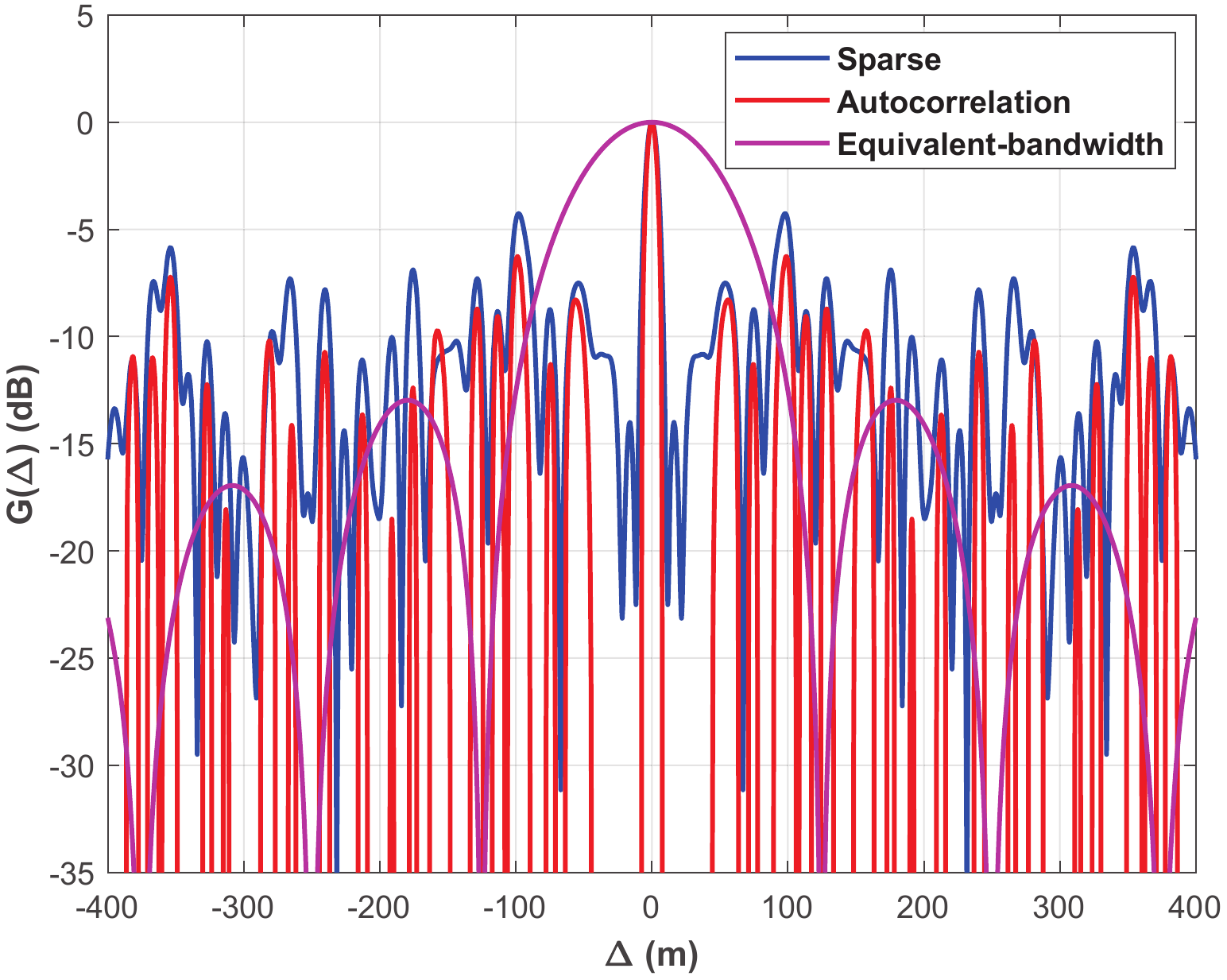}
    \caption{Ambiguity Function. }
    \label{F:AF}
\end{figure}

As shown in Fig. \ref{F:AF}, the autocorrelation-based method exhibits a better ambiguity function. Moreover, from \eqref{eq:26}, in the case of multiple OFDM symbols, the autocorrelation-based method can achieve better sensing performance.

The sensing signal on the virtual resource algorithm is summarized in Algorithm~\ref{alg:Sensing signal on virtual re}.

\begin{algorithm}[htbp]
\caption{Sensing Signal Generation on Virtual Resource}
\label{alg:Sensing signal on virtual re}
\begin{algorithmic}[1]
    \renewcommand{\algorithmicrequire}{\textbf{Input:}}
    \renewcommand{\algorithmicensure}{\textbf{Output:}}
\REQUIRE Subcarrier indices $\tilde{\mathcal{N}}$, received signal $\bm{Y}[n],n\in \tilde{N}$
\ENSURE Virtual resource element value $\tilde{\bm{R}}$, virtual resource element indices $\mathcal{P}$
\STATE Initialize $\bm{R} = \emptyset$, $\mathcal{P}=\emptyset$;
\STATE Take autocorrelation function of $\bm{Y}$, $\mathcal{R} =  \bm{Y}\bm{Y}^H$;
\STATE Compute virtual resource indices $\mathcal{D} = \tilde{\mathcal{N}} - \tilde{\mathcal{N}}^T$;
\STATE Get the virtual resource element value $\tilde{\bm{R}}_{m}$ by \eqref{Auto};
\STATE Coherently accumulate over OFDM symbols  $\tilde{\bm{R}}$ by \eqref{eq:26};
\STATE Get the virtual resource indices $\mathcal{P}$ by \eqref{eq:virtual signal indices};

\RETURN $\tilde{\bm{R}}$, $\mathcal{P}$;
\end{algorithmic}
\end{algorithm}

The delay estimation periodogram can be obtained by performing an IFFT operation to the sensing signal on the virtual resource, given by
\begin{align}
     &\mathcal{F}^{-1} \{\tilde{E\,[\bm{R}]}\}_q=\frac{1}{N}\sum_{s\in \mathcal{P}}(\sum_{l=1}^{L}|\alpha_l|^2 e^{-j 2\pi  \Delta fs\tau_l})e^{j 2 \pi \frac{q}{N}}
\end{align}

\section{Numerical Results}
In this section, numerical results are provided to evaluate the sensing performance of the proposed autocorrelation-based method. For the simulated ISAC system , the carrier frequency is 24 GHz, the number of subcarriers is 1000, and the subcarrier spacing is 120 kHz, with the conventional cyclic prefix (CP) ignored. The CPI is set to 6 ms in the simulation, and the number of OFDM symbols used is 720. The simulation settings are summarized in Table \ref{Tab:sim_params}. In the following simulation results, ``Sparse" represents the DFT-based method, ``Autocorrelation" represents the autocorrelation-based method, ``full-bandwidth" refers to using all available resources, and ``equivalent-bandwidth" refers to using an equivalent number of contiguous resources.
\begin{table}[h]
\centering
\caption{Parameter settings for simulation}
\label{Tab:sim_params}
\begin{tabular}{|l|l|}
\hline
 Symbols & Value \\
\hline
Carrier frequency  $f_c$ & 24 \rm{GHz}\\
Subcarrier spacing  $\Delta f$ & 120 \rm{kHz} \\
Number of subcarriers  $N$ & 1000 \\
Number of OFDM symbols  $M$ & 720 \\
Active subcarriers  $\tilde{N}$ & 200 \\
Full Bandwidth  $B$ & 122.88  \rm{MHz} \\
Transmit power  $P_t$ & 0.1  \rm{W} \\
Tx antenna gain  $G_{\text{tx}}$ & 20  \rm{dB} \\
Rx antenna gain  $G_{\text{rx}}$ & 20  \rm{dB} \\
\hline
\end{tabular}
\end{table}
\begin{figure}[htbp]
    \centering
    \includegraphics[scale=0.23]{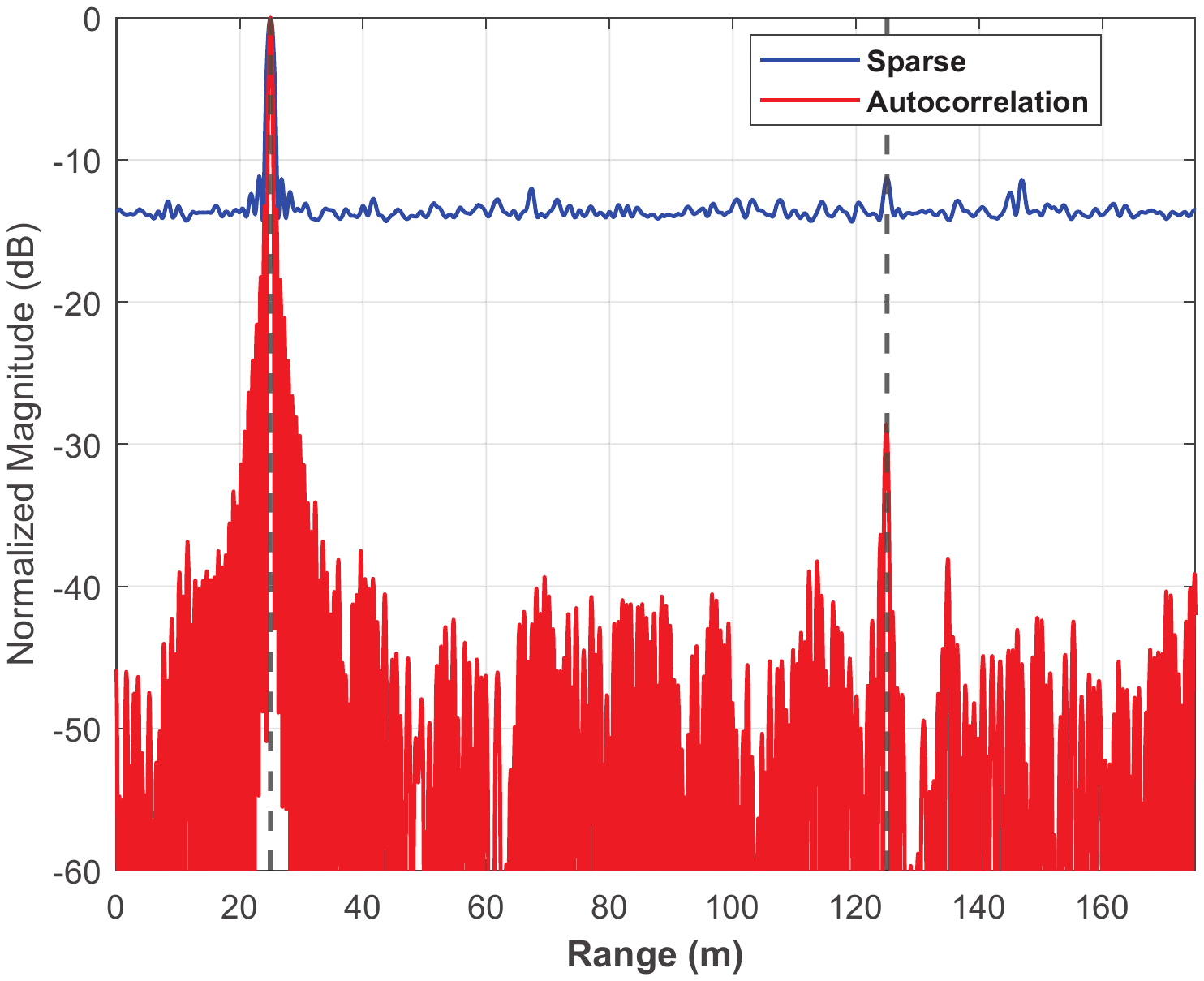}
    \caption{Multi-targets sensing results of the autocorrelation-based and DFT-based methods under sparse resource allocation at $\rm{SNR} = - 10 \rm{dB}$.}
    \label{F:mult sub1}
\end{figure}

Fig. \ref{F:mult sub1} shows the two-targets sensing results of the autocorrelation-based method and the DFT-based method with limited sparse resource allocation in $\rm{SNR} =- 10 \rm{dB}$. It can be observed that the autocorrelation-based method is capable of correctly detecting multiple targets due to its superior peak-to-sidelobe ratio (PSLR) characteristics.
\begin{figure}[htbp]
    \centering
    \includegraphics[scale=0.23]{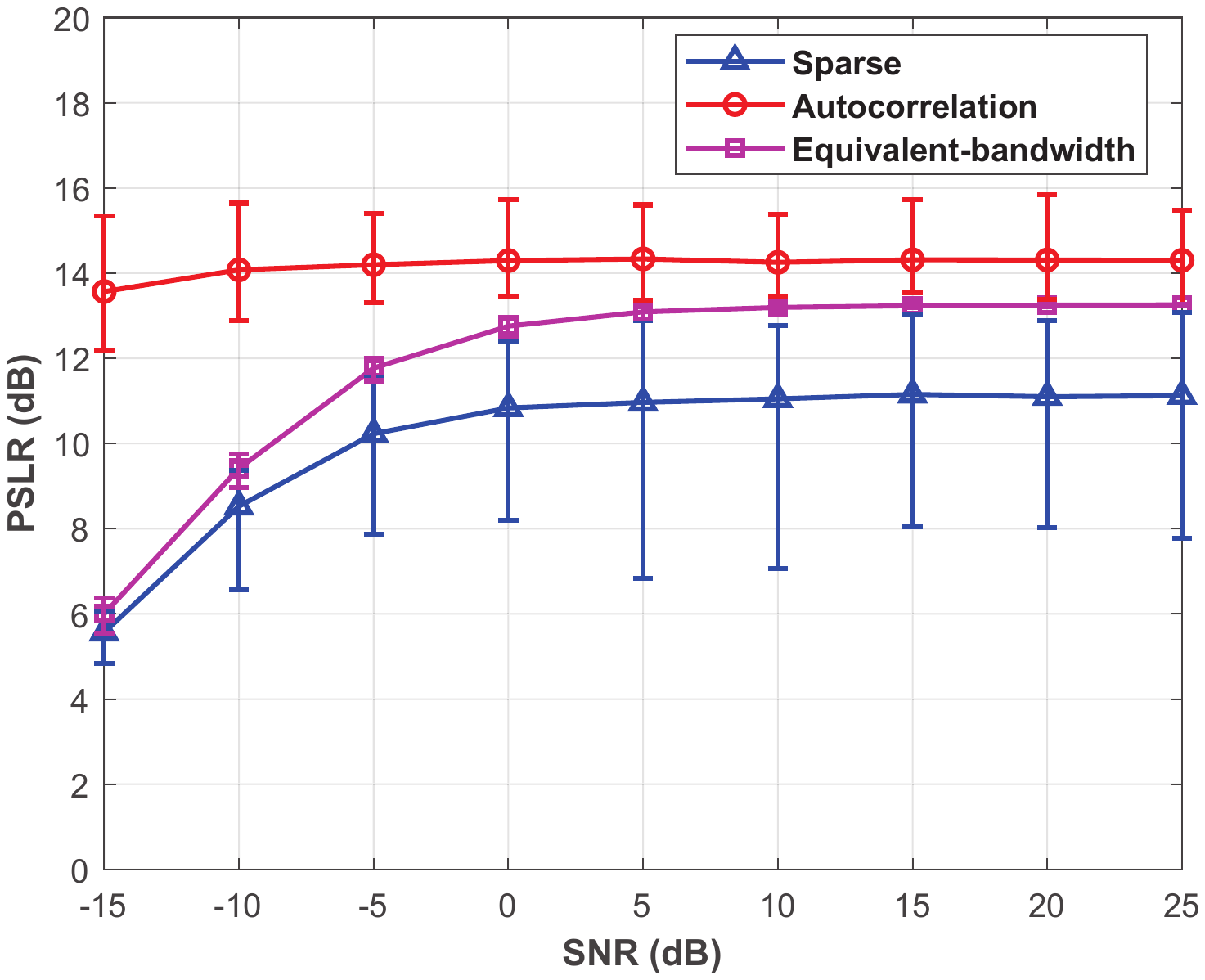}
    \caption{Average PSLR and 95\% confidence interval comparison under varying SNR conditions.}
    \label{F:PSLR}
\end{figure}

As shown in Fig.~\ref{F:PSLR}, the autocorrelation-based method achieves a higher PSLR and exhibits superior mainlobe-to-sidelobe characteristics compared to the DFT-based method and equivalent-bandwidth, resulting in enhanced noise robustness and sensing performance in multi-target scenarios. 
\begin{figure}[htbp]
    \centering
    \includegraphics[scale=0.23]{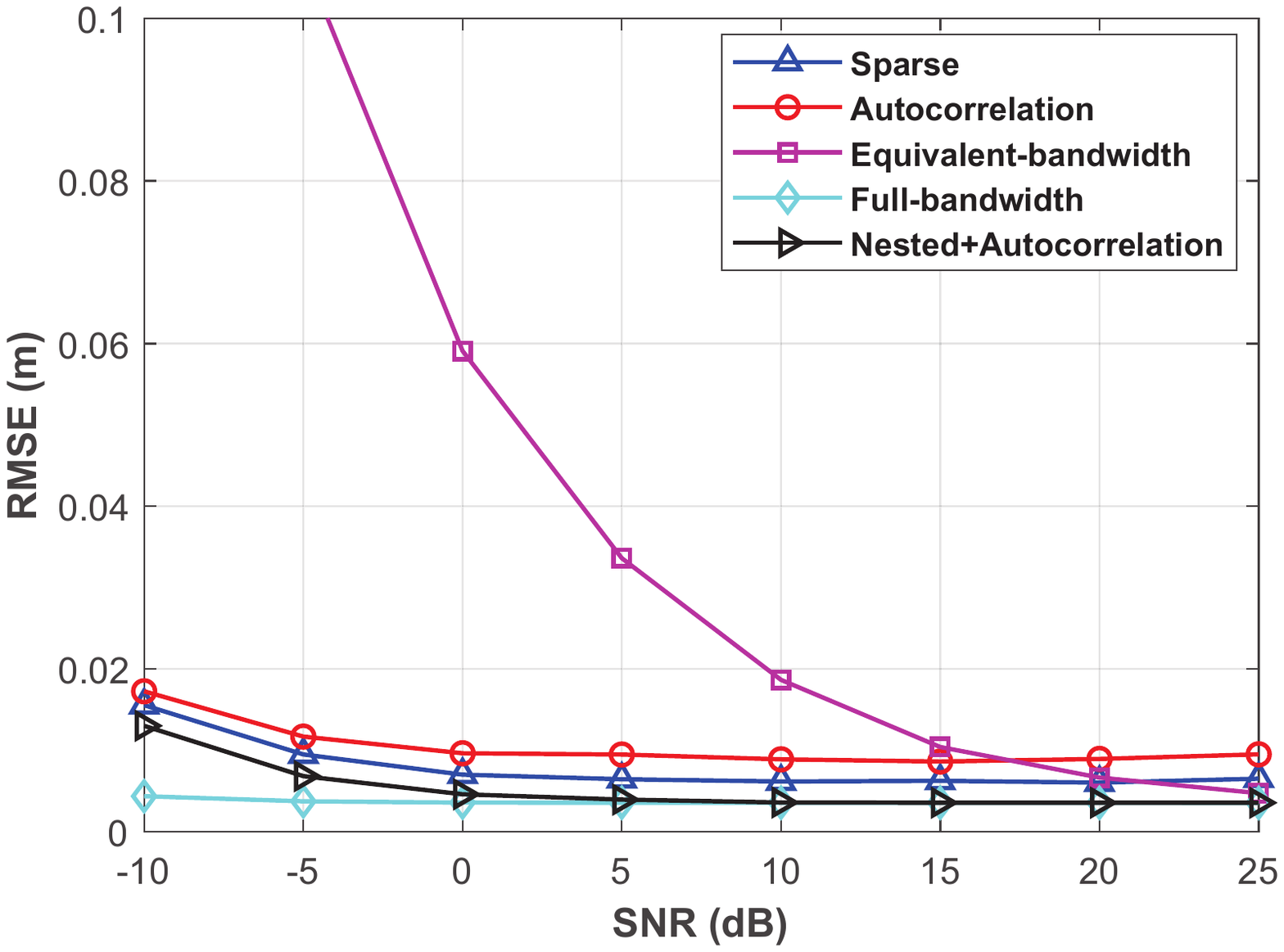}
    \caption{Range RMSE performance comparison under varying SNR conditions.}
    \label{F:RMSE}
\end{figure}

The nested resource allocation is a specific resource allocation suitable for the autocorrelation-based method. The nested resource allocation ensures coverage of all virtual resource positions, thus achieving better sensing performance, especially at high $\rm{SNR}$.

The proposed autocorrelation-based method achieves an RMSE close to that of the full-bandwidth resource allocation in range estimation, shown in Fig. \ref{F:RMSE}. Compared to the equivalent-bandwidth resource allocation, the DFT-based method and the proposed autocorrelation-based method achieve lower RMSE. However, the DFT-based method exhibits higher sidelobes than the proposed autocorrelation-based method, which makes it more susceptible to false alarms in multi-target scenarios.

\section{Conclusion}
In this paper, we analyzed the sensing parameter estimation for both single target and multiple targets in OFDM-ISAC system with non-uniform sparse resource allocation. The likelihood function and the CRLB of the parameter estimation are formulated when there is a single target.

Analysis confirms ML estimation is equivalent to conventional FFT with zero-filled unused resources. For multiple targets, our proposed autocorrelation-based method applies IFFT to virtual resources, notably enhancing PSLR while maintaining benchmark-comparable RMSE.

Excessive physical resources can introduce virtual redundancy and increase system load without effectively compensating for virtual source holes, leading to diminishing returns. Therefore, efficient strategies like minimum-redundancy resource allocation remain crucial for future high-performance designs.

\bibliographystyle{ieeetr}
\bibliography{ISACTutorial}

\end{document}